\documentstyle[eqsecnum,multicol,aps,prd,epsf]{revtex}
\newcommand{\smallfrac}[2] {\mbox{$\frac{#1}{#2}$}}
\newcommand {\eqref} [1] {(\ref {#1})}
\newcommand {\beq} {\begin{equation}} 
\newcommand {\eeq} {\end{equation}}
 \newcommand {\ber}{\begin{eqnarray*}}
 \newcommand {\eer} {\end{eqnarray*}}
\newcommand {\bea}{\begin{eqnarray}}
 \newcommand {\eea} {\end{eqnarray}} 
\newcommand{\Nfour} {${\cal N}=4\ $}
\newcommand{\Ntwo}{${\cal N}=2\ $}
\newcommand{\None}{${\cal N}=1\ $}

\def\a{\alpha}

\def\m{\mu}     \def\n{\nu}

\def\t{\tau}
\def\th{\theta} \def\Th{\theta}

\def\cJ{{\cal J}}  
\def\ii{{\rm i}}

\def\Acknowledgements{\bigskip  \bigskip {\begin{center} \begin{large}
             \bf ACKNOWLEDGEMENTS \end{large}\end{center}}}

\begin{document}
\rightline{{CPTH-S031.0601}}
\rightline{{LPTENS-01/30}}
\vskip 1cm
\centerline{{\Large \bf Non--Commutative Gauge Theories}}
\vskip 0.2cm
\centerline{{\Large \bf and the Cosmological Constant}}
\vskip 1cm
\centerline{Adi Armoni${}^a$ and Rodolfo Russo${}^b$}
\vskip 0.5cm
\centerline{${}^a$ Centre de Physique Th{\'e}orique de l'{\'E}cole 
Polytechnique~\footnote{Unit{\'e} mixte  du
CNRS et de  l'{\'E}cole Polytechnique,  
UMR 7644.}}
\centerline{F-91128 Palaiseau Cedex}
\vskip 0.3cm
\centerline{armoni@cpht.polytechnique.fr}
\vskip 0.5cm
\centerline{${}^b$ Laboratoire de Physique Th{\'e}orique de
l'{\'E}cole Normale Sup{\'e}rieure\footnote{Unit{\'e} mixte  du
CNRS et de l'{\'E}ole Normale Sup{\'e}rieure,  
UMR 8549.}}
\centerline{24, rue Lhomond, F-75231 Paris Cedex 05}
\vskip 0.3cm
\centerline{rodolfo.russo@lpt.ens.fr}
\vskip 0.3cm 

\begin{abstract}
We discuss the issue of the cosmological constant in non--commutative
 non-supersymmetric gauge theories. In particular, in orbifold field
 theories non--commutativity acts as a UV cut-off. We suggest that in
 these theories quantum corrections give rise to a vacuum energy $\rho$, that
 is controlled by the non--commutativity parameter $\th$,  $\rho \sim
 {1\over \theta ^2}$ (only a soft logarithmic dependence on the Planck
 scale survives). We demonstrate our claim in a
 two-loop computation in field theory and by certain higher loop
 examples.  Based on general expressions from string theory, we
 suggest that the vacuum energy is controlled by non--commutativity to
 all orders in perturbation theory.
\end{abstract}

\section{Introduction}
Explaining the small observed value of the cosmological constant is
 one of the crucial problems in contemporary theoretical physics
 (see~\cite{Weinberg:1989cp,Carroll:2001fy,Witten:2000zk} for
 reviews). Although
 classically the value of the cosmological constant can be tuned to
 an arbitrarily small value, one cannot do so in a quantum theory. In
 general quantum fluctuations force one to relate the vacuum energy
 density to the UV cut-off scale in the theory, $\rho \sim \Lambda^4$.
 From the theoretical point of view, the most natural thing is to
 identify $\Lambda$ with the Planck scale. However, this yields a
 cosmological constant which is enormously bigger than the estimated
 value.

In supersymmetric theories the vacuum energy 
 is exactly zero, due to the cancellation of fermionic and bosonic
 fluctuations. However, once supersymmetry is broken,
 ``mixed'' contributions
 of the form $\Lambda ^2 _P \sum (m^2 _B - m^2 _F)$ arise (where $m_B$ and
 $m_F$ are bosonic and fermionic masses respectively). 
 Thus, although supersymmetry might reduce the value
 of the vacuum energy, the problem remains unsolved.

It was suggested that certain non-supersymmetric theories,
 called ``orbifold field theories''~\cite{Kachru:1998ys},
 might lead to a
 small value of the vacuum energy. From the field theory 
 point of view the reason is the following: the planar graphs of
 these theories are exactly the same as the planar graphs of the 
 parent supersymmetric theories~\cite{Bershadsky:1998mb}. Therefore, at
 the large $N$ limit, the value of the vacuum energy is exactly zero.
 The problem is that there is no control on the non-planar graphs. In
 particular, $U(1)$ contributions lead to a large vacuum energy $\rho
 \sim {1\over N^2} \Lambda_ P ^4$.

In this note we combine the above mentioned approach with the
suggestion that non--commutativity might be a useful ingredient in the
solution of the cosmological constant
problem~\cite{Minwalla:2000px,Witten:2000zk}.  In fact,
non--commutativity can play the role of a UV cut--off for non-planar
graphs~\cite{Filk:1996dm}. In a generic non--commutative field theory
this fact would not improve the situation. The planar graphs would
contribute exactly as the ordinary theory, namely $\rho _{planar} \sim
\Lambda _P ^4$. However, in non--commutative orbifold theory planar
graphs are the same as in the parent supersymmetric
theory~\cite{Bershadsky:1998mb} and thus do not contribute to the
vacuum energy. The contributions from non-planar graphs are expected
to be controlled by the non--commutativity parameter. Thus, in this
setup, there is the possibility to disentangle the cosmological
constant problem from the value of the UV cut--off.

 As we shall see, the situation is a bit more involved than that. In
 fact, it is well known that the presence of time--like
 non--commutativity implies a breakdown of
 unitarity~\cite{Gomis:2000zz}, making the field theory not
 consistent. Thus we are forced to consider four dimensional field
 theories with only two non--commuting space directions
 ($[x^{1},x^{2}] = \ii\, \theta \neq 0$) . In this case the presence
 of $\theta$ {\em does not} completely regulate the loop integrals. In
 particular, UV divergences can appear in non--planar graph when the
 Moyal phases involve only internal ({\em i. e.} integrated)
 momenta. Of course this is always the case for the vacuum bubbles,
 where there are no external legs.  For example, the first
 contribution to the vacuum energy coming from a non-planar diagram is
 at two loop level. This contribution is, indeed, controlled by the
 non--commutativity scale, but in theories with only space-space
 non--commutativity a soft logarithmic dependence on the Planck scale
 survives. This behavior is not surprising: a logarithmic dependence
 on the UV cut-off is typical of two dimensional theories and, because
 of the presence of space-like $\theta$, also in our four dimensional
 theories UV divergences can come only from the commutative plane.
 
We suggest that this situation persists to all orders and that the
leading value of the vacuum energy in non-supersymmetric orbifold
field theories is 
\beq 
\rho \sim {1\over \theta ^2} P (\lambda,\log \sqrt \theta
\Lambda_P) \label{result} , 
\eeq
where $P$ is a polynomial of the 't~Hooft coupling 
$ \lambda = g^2 N$ and the $log$.
Note that the leading term of expression~\eqref{result} is also
suppressed by ${1\over N^2}$ with respect to the usual result, since
the vacuum energy generically is $\sim N^2$.

Some remarks are necessary in order to make this statement more
precise. First, eq.~\eqref{result} strongly depends on the UV/IR mixing and
on the possibility 
of re-summing classes of non-planar diagrams, as suggested
in~\cite{Minwalla:2000px}. However, it turns out that this re-summation
can not be performed when the number of bosonic degrees of freedom in
the adjoint representation exceeds that of the adjoint fermions. This
pathology was to be expected, since for these non--commutative field
theories the 1-loop dispersion relation signals that the trivial
vacuum is quantum--mechanically unstable~\cite{Ruiz:2001hu}. In
general, the nature of the vacuum state of the non--commutative
field theories has not been fully understood yet. However, in the
case where there are more fermionic fields in the adjoint
representation than bosonic ones, the re-summation proposed
by~\cite{Minwalla:2000px} can be performed without creating any
evident instability. Thus, for these theories, we will take a
pragmatic approach: we will consider the expansion around the usual
vacuum and analyze the generic degree of divergence of the vacuum
bubbles. 

In addition, in the usual commutative case vacuum bubbles are
immediately related to the cosmological constant because of Lorentz
invariance. On the contrary, in non--commutative theories, this
invariance is explicitly broken and the v.e.v. of the matter
energy--momentum tensor can take a more general form 
$\langle T_{\mu\nu}\rangle = - \rho \, g_{\mu\nu} + \sigma\,
(\theta^{-2})_{\mu\nu}$. In this setup vacuum diagrams correspond to
the combination of the above two terms. We will not separate the
contribution related to $\rho$, because we are interested in the
dependence on $\Lambda_P$, that is of the form of eq.~\eqref{result}
for both $\rho$ and $\sigma$. 

Moreover, there are ``phenomenological'' obstacles that generally make 
the relation between non--commutative theories and the real world
difficult. A realization of a
consistent quantum NC $SU(N)$ model is not at hand at the moment (see
however~\cite{Jurco:2001rq} for a recent attempt in this direction).
In particular, the baryon number is always gauged in non--commutative
theories. In addition, due to UV/IR mixing it is not clear whether a
non-supersymmetric non--commutative theory would flow in the IR to
its commutative counterpart. A recent analysis of the non-commutative
\Ntwo model suggests that this indeed might be the
case~\cite{Armoni:2001br,Hollowood:2001ng}. The running of the
effective coupling in non-commutative \None SQCD also supports such a scenario
\cite{Chu:2001fe}.  

Finally, the cosmological constant receives contributions also from the
gravity sector (closed strings) beyond those coming from the gauge theory
sector (open string). The non--commutativity we consider does not
affect gravitational interactions (closed string amplitudes) and thus
$\theta$ cannot act as a regulator for those contributions. 

The organization of this manuscript is as follows.
In Section~2 we briefly recall the main features of the
orbifold field theories derived from the supersymmetric gauge theories.
In Section~3 we focus on vacuum diagrams and argue that, in 
these theories, the non--commutative parameter plays a crucial role in
regulating the UV divergences.
In Section~4 we use the D-brane picture of the orbifold field theories
to discuss the general structure of the vacuum diagrams. 

\section{Orbifold field theories}

Orbifold field theories are obtained by a certain truncation of a
supersymmetric gauge theory. For instance, let us consider the special
case of \Nfour super\-symmetric Yang-Mills theory. The truncation
procedure is as follows: consider a discrete subgroup $\Gamma$ of the
\Nfour R-symmetry group $SU(4)$. For each element of the orbifold
group, the regular representation $\gamma$ inside $U(|\Gamma |N)$
should be specified. Each field $\Phi$ transform as $\Phi \rightarrow
r \gamma^\dagger \Phi \gamma$, where $r$ is a representation matrix
inside the R-symmetry group. The truncation is achieved by keeping
invariant fields.  The resulting theory has a reduced amount of
supersymmetry, or no supersymmetry at all. It was
suggested~\cite{Kachru:1998ys}, based on the AdS/CFT conjecture, that
the truncated large $N$ theories are finite as the parent \Nfour
theory. Later it was proved~\cite{Bershadsky:1998mb} that the planar
diagrams of the truncated theory and parent theories are identical.

Let us consider a specific example~\cite{Klebanov:1999ch}.  The
 example is an $U(N) \times U(N)$ gauge theory with 6 scalars in the
 adjoint of each of the gauge groups and 4 Weyl fermions in the
 $(N,\bar N)$ and 4 Weyl fermions in the $(\bar N,N)$ bi-fundamental
 representations. This is the theory that lives on dyonic D3 branes of
 type 0 string theory and can be also understood, from the field
 theory point of view, as a $Z_2$ orbifold projection of \Nfour
 SYM~\cite{Nekrasov:1999mn}. Other examples
 of large $N$ finite non-supersymmetric theories can be found
 in~\cite{Armoni:1999gc,Blumenhagen:1999uy}.

 In all these examples there are more bosonic fields in the adjoint
 representation than fermionic fields in the adjoint representation.
 As was already mentioned in the introduction, the non--commutative
 version of these field theories displays ``tachyonic''
 instabilities~\cite{Ruiz:2001hu} and is not useful for our purposes.
 Thus we wish to give an example of an orbifold field theory with more
 fermionic fields in the adjoint representation than bosonic
 ones. We will start from a $U(2 N)$ \None SYM with chiral multiplets
 and no superpotential as a parent theory, and perform a projection
 similar to the one just described to get a $U(N)\times U(N)$ gauge
 theory. The content of the ``vector  multiplet'' in the orbifolded
 theory is a vector in the adjoint of each of the group factor and
 a fermion in the bi-fundamental representation. 
 In addition we can have
  $F$ copies of ``chiral multiplets'' with fermions in the
 adjoint of each group and scalars in the bi-fundamental. We do not
 add a superpotential to the action. 
 The planar sector of this theory is in one--to--one correspondence with
 \None SYM with $F$ chiral multiplets. Note also that when $F>1$
 we have more fermions than bosons in the adjoint representation.
 
\section{Field theory analysis}

In this section we would like to see how a small vacuum energy
 can be achieved in a world with large non--commutativity.
 The trace of the v.e.v. of the energy--momentum tensor is related to
 the partition function as follows
\beq
 -\ii V \tau = \log Z,
\eeq
where $V$ is the volume of the system.

For a free field theory, it is possible to perform the Gaussian
integration and to calculate $\tau$. The bosons and fermions 
contribute with opposite signs. The result is
\beq
\tau^{(1)}= 2\,(N_{B} - N_{F})~N^2\; {1\over 2} \int  {d^4 k \over (2\pi)^4} 
\log k^2~,
\eeq
and can be viewed, diagrammatically, as a sum over configurations
of bosonic or fermionic loops.  $N_B$ and $N_F$ count the bosonic
and the fermionic degrees of freedom respectively. In the specific
case of the $U(N)\times U(N)$ theory related to \Nfour one
has $N_B=N_F=8$ (six scalars plus the two physical degrees of freedom
of the gluon for the bosons and two physical d.o.f. for each one of
the four Weyl fermions), while for the example related to \None one
has $N_B=N_F=2+2F$.  This pattern is general to all orbifold field 
theories: the number of bosons and fermions is the same and the
cosmological constant vanishes at the one loop level.

Let us turn now to the two loop calculation. The various contributions
are described in figure \eqref{two} below.
   
\begin{figure}[ht]
\begin{center}
\begin{tabular}{c}
\epsfxsize=8cm
\epsfbox{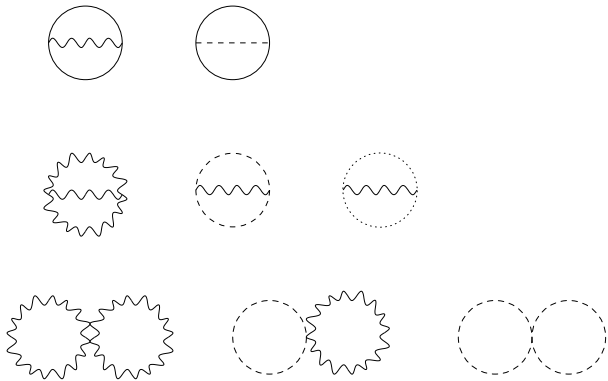}
\end{tabular}
\end{center}
\caption{Two loops contributions to the vacuum energy. Solid, wavy,
dashed and dotted lines represent spinors, gluons, scalars and ghosts
respectively.}
\label{two}
\end{figure}

In the parent non--commutative theory, the sum of all
contributions vanishes as required by supersymmetry.
In the case of the orbifold theory, the situation is different. Some 
fields are in the bi-fundamental representation whereas others are in
the adjoint and the Feynman rules for the two representations are
clearly different. For example: the vertex of a gluon coupled to
fundamental matter is 
\beq
 V_{fund} = g \times T ^a _{ij} \times \exp {(\smallfrac{\ii}{2}
p \theta q)} ,
\eeq
while for a gluon coupled to an adjoint matter the vertex is
\beq
 V_{adj} = {1\over 2} g \times \left( {\rm tr}\ [T^a,T^b]T^c \times
\cos{(\smallfrac{1}{2} p \theta q)} + \ii\;
{\rm tr}\ \{ T^a,T^b \} T^c \times 
\sin{(\smallfrac{1}{2} p \theta q)} \right) .
\eeq
For a full list of Feynman rules, in the case of adjoint matter
see~\cite{Armoni:2001xr,Bonora:2001ga}. 
Let us start by focusing on the planar graphs. The sum of these
contributions takes the following form  
\beq
\tau ^{(2)} _{planar} = \frac{\ii}{2}\;N_{B} \Big(N_{B} - N_{F}\Big)~ N^2 (\ii \lambda)
\int {d^4 p \over (2\pi)^4} {d^4 q \over (2\pi)^4} {-\ii\over p^2}
{-\ii\over q^2}~, \label{NC}
\eeq
The vacuum energy vanishes in the planar limit, as required by the
properties inherited from the original supersymmetric theory.
The contribution related to the non--planar diagrams breaks this
cancellation. In 't Hooft's two index notation, the reason for this
non-vanishing contribution is the presence of non-planar diagrams
which exist only for the case of fields in the adjoint representation.
 On the contrary, fields in the bi-fundamental representation cannot
give rise to any non-planar vertex.  Thus, the only 2-loop non-planar
diagrams come from the adjoint sector of the theory 
and the sum of these contributions is
\beq
\tau ^{(2)}_{non-planar} = - \frac{1}{2} \lambda~ N_B^{A} 
\,\Big(N_B^{A} - N _F^{A}\Big) 
\int  {d^4 p \over
(2\pi)^4} {d^4 q \over (2\pi)^4} {1\over p^2} {1\over q^2} 
\,{\rm e}^{\ii p\theta q},
\label{sum}
\eeq
where $N_B^{A}$ is the number of bosons in the adjoint representation
and $N_F^{A}$ is the number of fermions in adjoint representation.
In the specific example of the \Nfour daughter theory, $N_B^{A}=8$ and 
$N_F^{A}=0$ and hence $N_B^{A}-N_F^{A}>0$, whereas in the
example of the \None daughter theory $N_B^{A}=2, N_F^{A}=2F$ and
hence $N_B^{A}-N_F^{A}<0$ for $F>1$. As we shall see, this
difference in sign will be important later on. 
 
It is convenient (and useful for comparison with string theory
amplitudes) to use the Schwinger parameterization in order to calculate
\eqref{sum}
\beq
\tau ^{(2)} _{non-planar} = 
- \frac{1}{2} \frac{\lambda}{(4\pi)^4}~ N_B^{A}\,
\Big(N_B^{A}-N_F^{A}\Big) \int _0 ^\infty dt_1
 dt_2 {1\over t_1 t_2 (t_1 t_2 + \frac{\theta^2}{4})} \label{int}
\eeq

The integral \eqref{int} is both UV divergent (small $t$) and IR
divergent. The IR divergences are not manifest in \eqref{int}, but
exist due to massless degrees of freedom (see
\eqref{sum}). In order to regulate \eqref{int}, both UV and IR cut-offs
 are needed. In principle a regulator
which preserves gauge invariance should be used. However, since
we know that in the supersymmetric case the sum of all the contributions should
vanish, all the results reported in this work are determined
unambiguously by using a naive cut-off. 
Once the integral is regulated, the vacuum energy is
suppressed by ${1\over \theta ^2}$. Note also that the UV divergences are only
 logarithmic due to the presence of $\theta$. Let us ignore for a
moment the logarithmic IR divergences. The result is
\beq
\tau ^{(2)} \sim -\lambda\; {N_B^{A}(N_B^{A}-N_F^{A})\over \theta ^2 } \;
(\log \sqrt \theta \Lambda_P )^2 
\eeq 

\begin{figure}[ht]
\begin{center}
\begin{tabular}{c}
\epsfxsize=6cm
\epsfbox{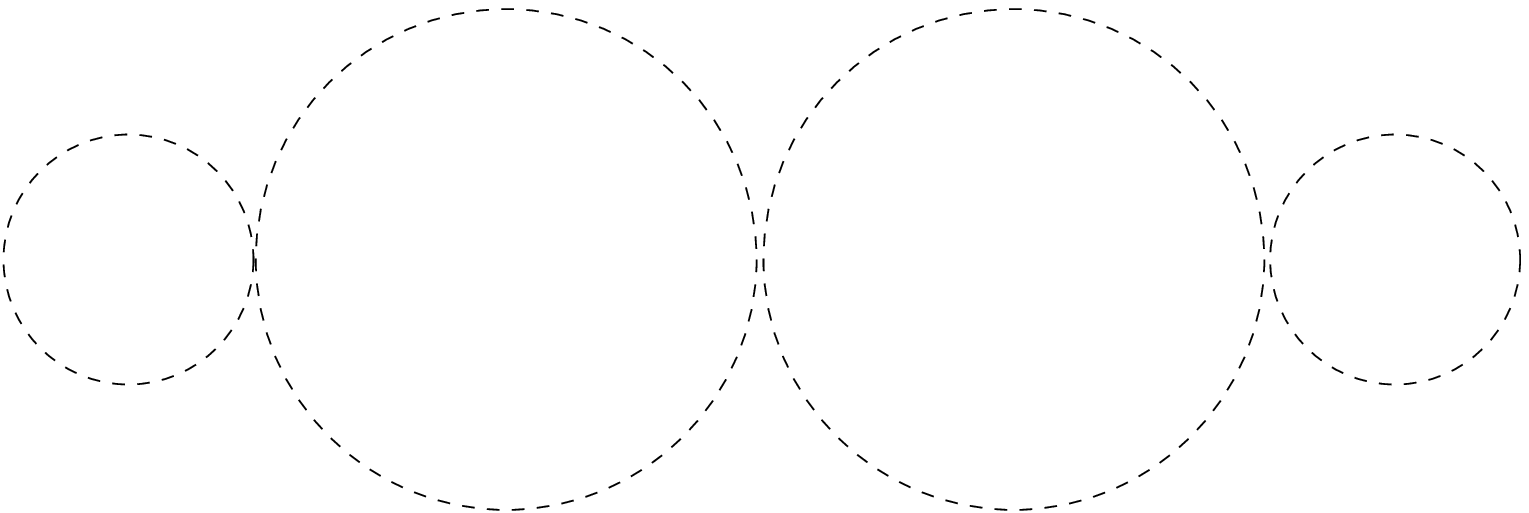}
\end{tabular}
\end{center}
\caption{A four loop example which involves three quartic vertices.}
\label{four}
\end{figure}

We would like to give another example. The example is a 4-loops
non-planar diagram which involves quartic interactions. Although we do
not consider all other 4-loop contributions (and even all 3-loop
contributions), we compute this diagram since it captures the nature
of diagrams with maximal non-planarity.  Maximal non-planarity means
that the determinant of the intersection matrix, describing which
propagators are crossing, is non-vanishing. In the next section we
will give a string picture of the field theory perturbation expansion
in the non--commutative case and it will be more clear why these
diagrams have a special status.  The non-planar 2-loop diagrams and
the 4-loop example given in figure \eqref{four} below are of this
type.

The corresponding expression is
\bea \label{4ex}
\tau ^{(4)} & \sim & 
\lambda\, g^4 \left (\int {d^4 p \over (2\pi)^4} {d^4 q \over (2\pi)^4}
\left ( {1\over p^2} \right ) ^2 {1\over q^2} {\rm e}^{\ii p\theta q}
\right )^2 \nonumber
\\ & = &
{1\over N^2} \frac{\lambda^3}{(4\pi)^{8}} 
\left( \int t_1 dt_1 dt_2 {1\over t_1 t_2
(t_1 t_2 + \frac{\theta^2}{4})} \right )^2,
\eea
which yields a contribution that is suppressed
by higher powers of $N$ and $\theta$ with respect to the two loop one
(figure \eqref{two}). In addition, due to UV/IR mixing effects,
there is a contribution to \eqref{4ex} of the form 
\beq
 \Lambda _P ^4 \left (\log
  (1+{1\over \theta ^2 m^2 \Lambda _P^2})\right )^2,
\eeq
which we can neglect. This is justified if we assume that $\theta\,
m\,\Lambda_P \gg 1$ (where $m$ is the IR cut-off) and, in this limit,
only negative power of $\Lambda_P$ are generated.

The general rule is that higher genus maximal
non-planar diagram are suppressed by both powers of $N^2$ and $\theta$.

So far we ignored the infra-red divergences appearing in the
integrals over Schwinger parameters. The reason is that these IR
divergences might disappear after re-summation of higher order
contributions, as suggested by~\cite{Minwalla:2000px}. 
Consider the non-planar diagram in figure \eqref{loops} below.

\begin{figure}[ht]
\begin{center}
\begin{tabular}{c}
\epsfxsize=4cm
\epsfbox{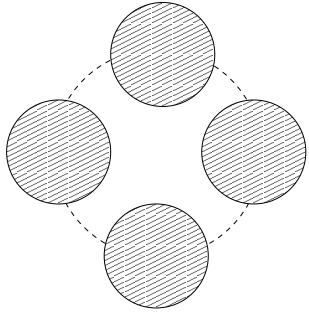}
\end{tabular}
\end{center}
\caption{Higher loop contribution to the vacuum energy. The dashed
lines represents scalars. The filled bubbles represent a loop with
 either bosons or fermions.}
\label{loops}
\end{figure}

This is a typical higher loop diagram. Denoting the internal momenta
in the dashed loops by $q_i$ and the momentum in the ``overall'' loop
by $p$, the potentially divergent contribution takes the form 
\beq
\left ( ( N_B^{A}- N_F^{A}) \lambda \right ) ^n \int {d^4 p\over
(2\pi)^4} {d^4 q_1 \over (2\pi)^4} {d^4 q_2 \over (2\pi)^4} ... {d^4
q_n\over (2\pi)^4} {\left ({1\over p^2}\right )}^n {1\over q^2 _1}
{1\over q^2 _2} ... {1\over q^2 _n} {\rm e}^{\ii p\theta \sum _i q_i}
\label{nloops} 
\eeq 
Upon integration over $q_i$, \eqref{nloops} takes the following form
\bea 
& & \left ( ( N_B^{A}- N_F^{A} )\lambda \right
) ^n \int {d^4 p \over (2\pi)^4} {\left ({1\over p^2 (\theta
p)^2}\right )}^n \sim \nonumber \\ & & \left ( (N_B^{A} - N_F^{A})
\lambda \right )^n \int dt_1 dt_2 \left ( {t_1\over t_2} \right )
^{n-1} {1\over t_1 t_2 (t_1 t_2 + \frac{\theta^2}{4})}
\label{ordern}.
\eea
At any order $n$ the contribution \eqref{ordern} seems to be
severely IR and UV divergent divergent.
 However, before integrating over $t_1,t_2$ let us
re-sum all orders. This is possible only when $N_F^{A}>N_B^{A}$,
 since only in this case does the series have alternating signs,
allowing us to arrive to a 
meaningful finite answer. Otherwise we will obtain a divergent answer.
 The result is
\beq
\sum _{n=1} ^\infty
((N_B^{A}-N_F^{A})\lambda)^n \int dt_1 dt_2 \left ( {t_1\over t_2}
\right ) ^{n-1} {1\over t_1 t_2 (t_1 t_2 + \frac{\theta ^2}{4})}
\sim -\lambda\; {1\over \theta^2} \log(\sqrt \theta \Lambda_P) \label{all} 
\eeq
 The integral is \eqref{all} logarithmically divergent. It suggests that upon
summation all orders the vacuum energy is indeed \eqref{result}.

Following \cite{Minwalla:2000px}, we suggest that the scenario that
was mentioned above is general: in each case where non-planarity
introduces IR divergences, these divergences would disappear upon
re-summation (or would result in at most as logarithmic IR
divergences). The algorithm is the following: each non-planar graph
can be ``opened'' by cutting a propagator, then it can be glued to
itself $n-1$ times and finally it can be ``closed'' by gluing the ends
of the propagator and integrating over the momentum flow. A
re-summation of all orders graphs should result with at most
logarithmic divergences. This procedure makes sense only in theories
with $N_F^{A}>N_B^{A}$.

Another important issue is the existence of mixed planar and
non-planar graphs. There is a danger that in such a case there will be
quadratic UV divergences. However, since the parent theory is
supersymmetric, this problem is avoided. The reason is that a generic
mixed diagram can be divided to its planar and non-planar parts connected
by propagators. Let us assume, for simplicity, that the propagators
are bosonic. Each propagator carries a factor of ${1\over p^2}$.
The planar part must be $\sim p^2 \log p/ \Lambda _P$,
since this sector is supersymmetric and therefore diverges, at most,
logarithmically. It means that the two propagators and the planar
piece together contributes $\sim 1/p^2 \log p/ \Lambda _P$. Therefore
they can be substituted by a single propagator (for the sake of
finding the $\theta$ dependence of the whole graph). It means that a non-planar
graph connected to a planar piece behaves as a fully non-planar graph.
This argument holds also for planar and non-planar graphs which are
connected by fermionic propagators or by ghosts. In the case of
fermionic propagators, the planar part of the graph diverges
logarithmically due to supersymmetry. In the case of ghosts, the 
reasoning is different: the planar graph must diverge at most 
logarithmically in order to preserve gauge invariance.

\section{String theory calculations}

It is well known that non--commutative field theories can be nicely
embedded in string theory~\cite{Seiberg:1999vs}. This means that, also
in the non--commutative case, it is possible to perform a particular
decoupling limit on various string quantities and recover the
corresponding field theoretic results.  Indeed the string setup seems
to be the most natural one for studying non--commutativity.  In fact,
at a field theory level the presence of non--commutativity requires to
modify the form of the interactions terms in the microscopic Lagrangian.
In a string setup, on the contrary, the building blocks, like the
$3$-string vertex operator, are unmodified~\cite{Chu:2000wp}, and, in
order to recover non--commutativity, it is sufficient to expand the
usual theory around a slightly different vacuum containing a constant
$B$-field. In the decoupling limit ($\alpha'\to 0$), a particular
combination of $B$ and $\alpha'$ is kept fixed and gives rise to the
non--commutative parameter $\theta_{\mu\nu}$. 

From the world-sheet point of view the presence of $B$
just affects the commutations relations among the string
modes~\cite{Chu:1999qz} and the nontrivial modification is concentrated
in the zero-modes part. Because of this, it was possible to
derive~\cite{Chu:2000wp}, at least in the bosonic case, a master
formula which contains any loop string amplitude in presence of a
background $B$-field. The same decoupling limit used at tree level can
be applied on this formula and many explicit
checks~\cite{Andreev:2000rm,Chu:2000wp} showed that in this way one
can recover {\em exactly} the Feynman diagrams of various field
theories.
Thus, the embedding of non--commutative field theory in a string setup
is not only useful at a conceptual level, but also represents a
simplifying technique for the computation of perturbative
amplitudes. This is the same pattern that emerged after the first
string revolution in the study of the usual gauge and gravity
interactions~\cite{a0}.

\begin{figure}[ht]
\begin{center}
\begin{tabular}{c}
\epsfxsize=4cm
\epsfbox{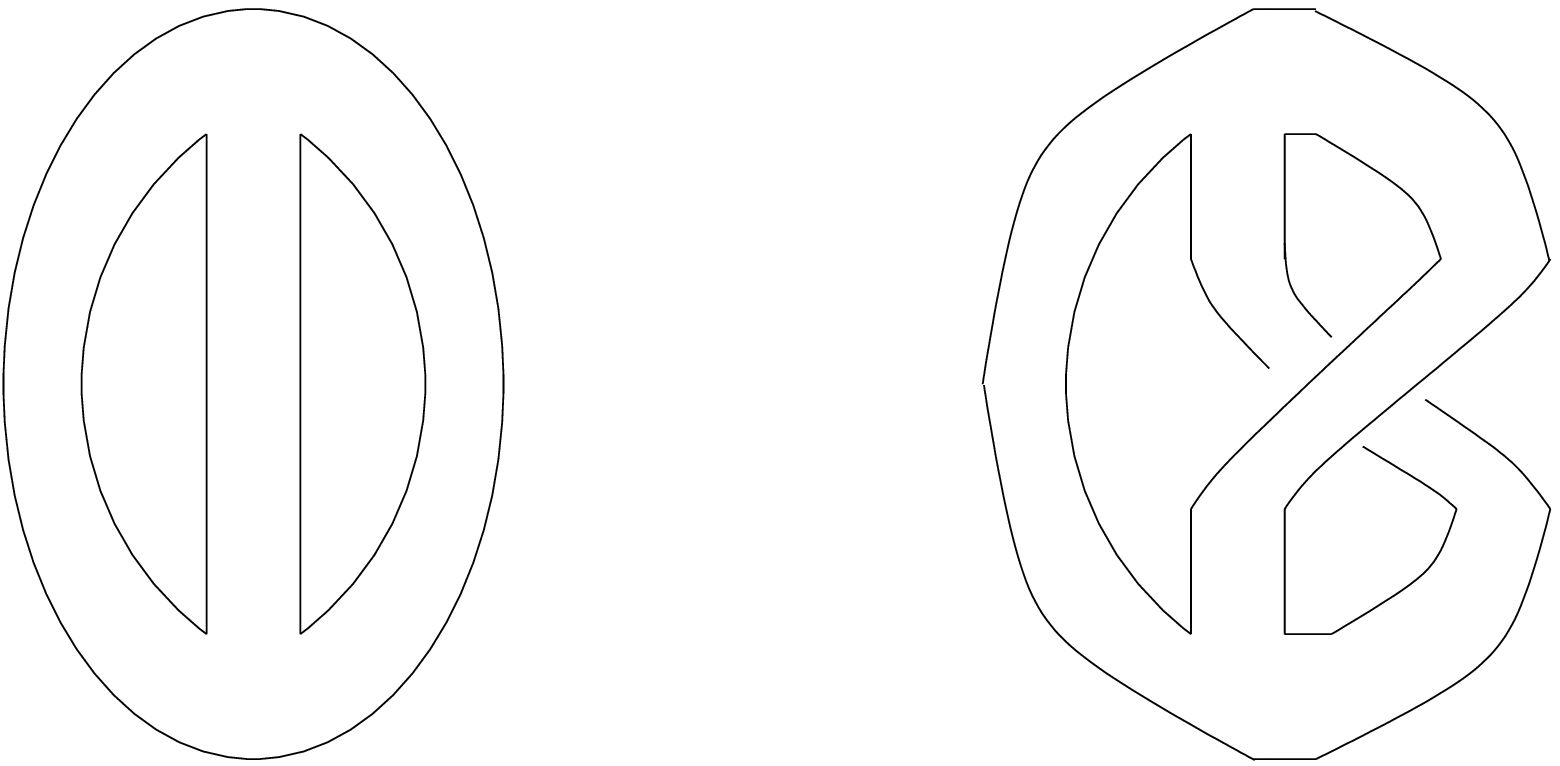}
\end{tabular}
\end{center}
\caption{The oriented two--loops surfaces. The one at the left has 3
borders, while the one at the right has only one border.}
\label{2ld}
\end{figure}

In this section we will use the general result derived at the string
level to study the main features of the higher loops vacuum bubbles.
Even if the string formula we will use takes into account only bosonic
degrees of freedom, it is able to capture the relevant properties of
the non--planar diagrams. 
At each order in the string perturbative expansion, all the vacuum
graphs are re-summed in few Riemann surfaces. Remembering that we need
to consider only oriented open strings, at the first order we
encounter just the annulus which, in the $\alpha'\to 0$ limit,
degenerates into a planar diagram. At the next order we have two
possibilities: a planar disk with two holes and a non--planar surface
where two propagators have to cross, see Figure~(\ref{2ld}).

In general the string result depends on two topological properties of
the surfaces: the number of loops $h$ (or the number of independent
momenta flowing in the propagators) and the number of borders (or
equivalently the intersection matrix $\cJ_{IJ}$ among the internal
momenta). In a compact form the contribution to the vacuum bubbles can
be written as
\begin{equation} \label{stringmaster} 
V_{0;h}^\Th = \left[\sqrt{\det M}\right]^{1-h}
\int \left[\det \left(\frac{-A}{2}\right)
\right]^{-1/2} [dm]_h~,
\end{equation} 
where the integration is taken over the moduli of the Riemann surface
and $M$ is the open string metric $M_{\m\n} = g_{\m\n} - (B g^{-1}
B)_{\m\n}$. $A$ contains the period matrix $\t_{IJ}$, the
non--commutativity parameter $\Theta^{\m\n} = 2\pi\a' \Big( \frac{1}{g+ B} 
B \frac{1}{g- B}\Big)^{\mu\nu}$ and the intersection matrix
$\cJ_{IJ}$
\begin{equation}
A_{IJ}^{\m\n} = 2\a'(2\pi \ii \t_{IJ}) (M^{-1})^{\m\n}
- \ii \Th^{\m\n}\cJ_{IJ}~.
\label{abc}
\end{equation}
In order to make the comparison with the field theory result simpler,
one can rescale the string coordinates and simply use $\eta^{\mu\nu}$
as open string metric and $\th^{\mu\nu} = 2\pi\a' B^{\mu\nu}$ as
noncommutative parameter.
It is important to stress that, in \eqref{stringmaster}, the
determinant is taken over the space of both Lorentz $(\mu)$ and
loop indices $(I)$.
Just to give a flavor of how this expression can be made explicit we
report the expression for the measure in the case of bosonic string
written in the Schottky parameterization of the world--sheet surface
and refer to~\cite{DiVecchia:1989cy} for the derivation and the
explicit form of the other quantities
\beq 
\label{hmeasure} [dm]_h = \frac{1}{dV_{abc}} \prod_{I=1}^{h}
\left[ \frac{dk_I d \xi_I d \eta_I}{k_I^2 (\xi_I - \eta_I)^2} ( 1- k_I
)^2 \right] \prod_{\alpha}\,\!' \left[ \frac{\prod_{n=1}^{\infty} ( 1
- k_{\alpha}^{n})^{-d+2}}{ ( 1 - k_{\alpha})^{2}} \right]~.  
\eeq 

For each hole one needs to introduce 3 real parameters
$(\eta_I,\xi_I,k_I)$ that specify its position and its width. The
$k_\a$'s are the multipliers of the other elements of the Schottky
group and can be written as function of the elementary parameters
$(\eta_I,\xi_I,k_I)$, related to the generators of the group. Finally
$dV_{abc}$ remembers that, because of the projective invariance, one
has to fix the values of three of these variables. Thus at $h$ loops,
one has $3h-3$ integrations which is exactly the number of propagator
one has in a $\phi^3$--like vacuum bubble. In fact for each
propagator one can introduce a Schwinger parameter and the
moduli of the Riemann surface are generically connected to these
parameters by relations like $f(\eta_I,\xi_I,k_I) \sim
\exp{(-t/\alpha')}$. The exact form of such relations depends on
the specific corner of the moduli space one is looking at. The
contribution of each corner resum the results of the Feynman diagrams
with the same topology. This one-to-one mapping between string and
field theoretic result is pictorially intuitive. For instance, it is
possible to recover also diagrams with quartic interactions, both in
the usual~\cite{Marotta:2000re} and in the non--commutative
case~\cite{Bilal:2000bk}, when some of the string moduli are related
to Schwinger parameters of finite length in unit of $\alpha'$, which
means they are vanishing in the field theory limit.  Actually, the
case of diagrams with only four--point vertices is the easiest,
because all the factors of $\alpha'$ in the string formula cancel
after the change of variable from the string moduli to the Schwinger
parameter is performed~\cite{Frizzo:2001ez}. Thus it is sufficient to
consider only the leading order in $\alpha'$ in all the string
expressions appearing in \eqref{stringmaster}.

For instance, let see how the diagrams in Figure (\ref{2ld}), reduces
to the field theory diagrams of Figure (\ref{two}). In this case $A$
is a $2\times 2$ matrix with respect to the loop indices. In the
$SL(2,R)$ gauge $\xi_1=\infty$,  $\eta_1=0$ and $\xi_2=1$, the 
leading of the matrix $A$ entering in the non--planar 
diagram is
\beq \label{basicA}
A = 2\a'
\left(
\begin{array}{cc}
 -\eta^{\mu\nu} \, \log k_1  & \eta^{\mu\nu}\, \log\eta_2 + 
 \ii \frac{\th^{\mu\nu}}{2\a'} \cr 
 \eta^{\mu\nu}\, \log\eta_2 -
 \ii \frac{\th^{\mu\nu}}{2\a'} & 
 \eta^{\mu\nu}\, \log k_2 
\end{array}
\right) + \ldots 
\eeq
while for the planar diagram, one get a similar result, but without
$\th_{\m\n}$ terms in the  off-diagonal entries since the intersection
matrix is zero. In the case of quartic interactions, the propagator in
the middle, related to $\log\eta_2$, has vanishing length $\a'
\log\eta_2 \sim 0$, while $-\a' \log k_i=t_i $. 
In particular, for the non--planar bubbles that correspond to a surface
with $h=2$, but with only one border ($b=1$), one gets
\beq
V_{h=2,b=1} \sim \int dt_1 dt_2 \;\frac{1}{t_1 t_2 (t_1 t_2 +
\frac{\th^2}{4})}~. 
\eeq
As we noticed, this integral has at most logarithmic UV divergences. In
fact, the UV behavior is encoded in the region of small $t$'s. The
presence of the space--space non--commutative parameter is sufficient to
change the usual dependence on the Planck scale $\Lambda^4_P$ into a
milder $1/\th^2 \log{\sqrt \theta \Lambda _P}$. This pattern generalizes to
all the maximal non--planar diagrams, that is those with arbitrary $h$,
but always with one border (this request implies that $h$ is even). In
fact, for quartic interactions one gets
\beq\label{gen4}
V_{h,b=1} \sim \int\prod_{i=1}^{2h-2} dt_i 
\;\frac{1}{P_0^{(h)}(t_i) \left( \sum_{j=0}^{h/2} 
(\frac{\th^2}{4})^j ~ P_j^{(h-2j)}(t_i)
\right)}~, 
\eeq
where $P_j^{(a)}$ are polynomials of order $a$ in the Schwinger
parameters and are linear in each $t_i$. Here
$P^{(0)}(t_i)=1$. Thus, the presence of a term$\;\sim\th^h$
regulates the contribution coming from the Gaussian integration over
the plane where the non--commutativity is present. The other two
directions give the usual result, but again the UV divergences are
of logarithmic type $\sim dt_i/t_i$.

The presence of three--point vertices makes it more difficult to write
a general formula for the vacuum bubbles and to read from it the
singular behaviour for small $t$'s. In fact, in these diagrams there
are more propagators in comparison to the case of four--point vertices
and, thus, more Schwinger parameters. For instance, at two loops, one
needs to associate a Schwinger parameter also to $\eta_2$ ($-\a' \log
k_i=t_i + t_3$ and $-\a' \log\eta_2=t_3 $). This means that the change
of variable from the string moduli to the field theory $t$'s generates
extra factors of $1/\a'$ that need to be compensated by considering
also the subleading terms in the string expression of $1/\sqrt{\det A}$.
This expansion has two effects: it generates higher powers of the
polynomial appearing in the denominator of \eqref{gen4} and a new
polynomial at the numerator, that takes into account the dependence of
the three-point vertices on the inflowing momenta. For instance, a
generic $4$--loop diagram with six three--point vertices will look like 
\beq\label{ex5}
V_{4,b=1} \sim \int\prod_{i=1}^{9} dt_i 
\;\frac{Q^{(9)}(t_i)}
{\Big[P_0^{(4)}(t_i)\Big]^{5-x} \Big[\left( \sum_{j=0}^{2} 
(\frac{\th^2}{4})^j ~ P_j^{(4-2j)}(t_i)
\right)\Big]^x}~, 
\eeq
where $1\leq x \leq 4$. It is clear that the most divergent
contributions come from the terms with $x=1$. We explicitly computed
some diagrams of this kind and always found that $t$'s present in the
polynomial $Q^{(9)}$ cancel the power-like divergence of the
denominator. We think that this pattern is general, because the
presence of $\Lambda^2_P$ terms in the non--commutative computation
would imply the existence of an unwanted $\Lambda^6_P$ contribution in
the commutative case. 

What we said up to now is valid for the maximal non--planar diagrams,
that have only one border. From the string master formula it is clear
that the diagrams with $b>1$ are more divergent. In this case the
determinant of the intersection matrix is vanishing and thus the
regulating term in~\eqref{gen4} with $\th^h$ and no $t_i's$ is
absent. In general, the maximal power of $\th$ present is equal to the
rank of the first minor with nonvanishing determinant. The diagram
considered in Figure \eqref{loops} is of this kind, like all those
entering in the resummation procedure outlined in the previous
section, after the first one that is maximal non--planar. As we
already suggested, after resummation these power--like divergences
should disappear.  Clearly other diagrams with more than one border
are those containing a planar part. In this case it is clear that the
bosonic contribution will contain a power--like dependence on
$\Lambda_P$; however, as we argued in the previous section, these
terms disappear once also the fermion diagrams are taken into account,
since the planar part of the whole diagram is supersymmetric.

\Acknowledgements
We would like to thank C. Angelantonj, I. Antoniadis, C. Bachas, C.-S. Chu, E.
Dudas, E. Fuchs, C. Kounnas, E. Lopez and N. Nekrasov for discussions.
This work has been supported by the European Union under RTN contracts
HPRN-CT-2000-00122 and HPRN-CT-2000-00131.

\end{document}